\documentclass[fleqn,11pt]{article}

\usepackage{amsmath,amssymb,amsfonts,graphicx}

\usepackage{amsfonts,amssymb,cite}
\usepackage{graphicx}

\def\onecol{\onecolumn \mathindent 2em}

\def\noi{\noindent}

\newcommand{\Title}[1]{\noi {{\Large\bf #1}}\\[1ex]}

\def\Aunames#1{\noi{\bf #1}}
\def\au#1{${}^{#1}$}
\def\Addresses#1{\medskip\noi \protect
	\begin{description}\itemsep -3pt {\it #1} \end{description}}
\def\adr#1#2{\item[${}^{#1}$]{\it #2}}

\newcommand{\Abstract}[1]{\vskip 2mm \begin{center}
        \parbox{16.4cm}{\small\noi #1} \end{center}\medskip}

\def\email#1#2{\footnotetext[#1]{e-mail: #2}\addtocounter{footnote}{1}}

\def\nq{\hspace*{-1em}}
\def\nqq{\hspace*{-2em}}
\def\nhq{\hspace*{-0.5em}}

\def\Acknow#1{\subsection*{Acknowledgments} #1}

\usepackage{color}

\def\Jl#1#2{#1 {\bf #2},\ }

\def\ApJ#1 {\Jl{Astroph. J.}{#1}}
\def\CQG#1 {\Jl{Class. Quantum Grav.}{#1}}
\def\DAN#1 {\Jl{Dokl. AN SSSR}{#1}}
\def\GC#1 {\Jl{Grav. Cosmol.}{#1}}
\def\GRG#1 {\Jl{Gen. Rel. Grav.}{#1}}
\def\IJMPD#1 {\Jl{Int. J. Mod. Phys. D}{#1}}
\def\JETF#1 {\Jl{Zh. Eksp. Teor. Fiz.}{#1}}
\def\JETP#1 {\Jl{Sov. Phys. JETP}{#1}}
\def\JHEP#1 {\Jl{JHEP}{#1}}
\def\JMP#1 {\Jl{J. Math. Phys.}{#1}}
\def\NPB#1 {\Jl{Nucl. Phys. B}{#1}}
\def\NP#1 {\Jl{Nucl. Phys.}{#1}}
\def\PLA#1 {\Jl{Phys. Lett. A}{#1}}
\def\PLB#1 {\Jl{Phys. Lett. B}{#1}}
\def\PRD#1 {\Jl{Phys. Rev. D}{#1}}
\def\PRL#1 {\Jl{Phys. Rev. Lett.}{#1}}

\def\al{&\nhq}
\def\lal{&&\nqq {}}
\def\eq{Eq.\,}

\def\beq{\begin{equation}}
\def\eeq{\end{equation}}
\def\bear{\begin{eqnarray}}
\def\bearr{\begin{eqnarray} \lal}
\def\ear{\end{eqnarray}}
\def\earn{\nonumber \end{eqnarray}}
\def\nn{\nonumber\\ {}}

\def\nnn{\nonumber\\ \lal }
\def\nnnv{\nonumber\\[5pt] \lal }

\def\yyy{\\[5pt] \lal }
\def\eql{\al =\al}

\def\e{{\,\rm e}}
\def\d{\partial}

\def\sign{\mathop{\rm sign}\nolimits}
\def\diag{\mathop{\rm diag}\nolimits}

\def\const{{\rm const}}

\def\then{\ \Rightarrow\ }

\def\kappa{\varkappa}

\def\eqn#1{\eq\eqref{#1}}
\def\rf{\eqref}

\def\mN{_\mu^\nu}

\def\N{{\mathbb N}}

\def\cK{{\cal K}}

\def\sph{spherically symmetric}
\def\ssph{static, spherically symmetric}
\def\Scw{Schwarz\-schild}
\def\RN{Reiss\-ner--Nord\-str\"om}

\def\bh{black hole}
\def\bhs{black holes}
\def\wh{wormhole}
\def\whs{wormholes}
\def\asflat{asymptotically flat} 
\def\emag{electromagnetic}

\usepackage{color}

\begin{document}
\onecol
\thispagestyle{empty}

\Title{Magnetized dusty black holes and wormholes}

\Aunames{Kirill A. Bronnikov,\au{a,b,c,1} Pavel E. Kashargin,\au{d,2} Sergey V. Sushkov\au{d,3}}

\Addresses{\small
\adr a {Center of Gravitation and Fundamental Metrology, VNIIMS, 
		Ozyornaya St. 46, Moscow 119361, Russia} 
\adr b  {Peoples' Friendship University of Russia (RUDN University), 
		6 Miklukho-Maklaya St, Moscow, 117198, Russia}
\adr c  {National Research Nuclear University ``MEPhI'', 
		Kashirskoe sh. 31, Moscow 115409, Russia}
\adr d {Institute of Physics, Kazan Federal University, 
		Kremliovskaya St. 16a, Kazan 420008, Russia}
		}

\Abstract{We consider the generalized Tolman solution of general relativity, describing the evolution 
	of a spherical dust cloud in the presence of an external electric or magnetic field. The solution contains 
	three arbitrary functions $f(R)$, $F(R)$ and $\tau_0(R)$, where $R$ is a radial coordinate in the 
	comoving reference frame. The solution splits into three branches corresponding to hyperbolic ($f >0$),
	parabolic ($f=0$) and elliptic ($f < 0$) types of motion. In such models, we study the possible existence 
	of wormhole throats defined as spheres of minimum radius at a fixed time instant, and prove the existence 
	of throats in the elliptic branch under certain conditions imposed on the arbitrary functions. It is further 
	shown that the normal to a throat is a timelike vector (except for the instant of maximum expansion, when
	this vector is null), hence a throat is in general located in a T-region of space-time. Thus if such a dust 
	cloud is placed between two empty (\RN\ or Schwarzschild) space-time regions, the whole 
	configuration is a black hole rather than a wormhole. However, dust clouds with throats can be inscribed 
	into closed isotropic cosmological models filled with dust to form wormholes which exist for a finite 
	period of time and experience expansion and contraction together with the corresponding cosmology. 
	Explicit examples and numerical estimates are presented. The possible traversability of wormhole-like
	evolving dust layers is established by a numerical study of radial null geodesics. 
	}

\email 1 {kb20@yandex.ru}
\email 2 {pkashargin@mail.ru}
\email 3 {sergey\_sushkov@mail.ru}

\section{Introduction}

  A wormhole is a kind of spatial geometry resembling a tunnel that connects two different regions 
  of the same space-time, or two different space-times. Such geometries as solutions 
  to the gravitational field equations were first mentioned in 
  \cite{Flamm, EnstRos, Wheeler1, Wheeler2}, 
  but they were not traversable in the sense that a subluminal particle could not travel from one ``end
  of the tunnel'' to the other and return back. Probably the first exact traversable wormhole solutions 
  were discussed in \cite{Bronnikov1, Ellis} (1973) in the Einstein-scalar theory in which the scalar is of 
  phantom nature, that is, has a wrong sign of the kinetic term in the Lagrangian. Later, an evolving 
  version of this scalar-vacuum solution was found \cite{Ellis2}. One can also mention G. Clement's 
  papers \cite{Cle1, Cle2} describing static higher-dimensional \whs. But the greatest interest in 
  these objects arose after the work of M. Morris and K. Thorne \cite{MorTho} (1988) who 
  showed that, in the framework of Einstein gravity, maintaining a static wormhole throat 
  needs ``exotic'' matter that violates the Null Energy Condition (NEC). Overviews of wormhole
  research can be found, for example, in \cite{Visser, LoboRew}. 
  
  Wormholes have been considered in various theories of gravity and diverse models of matter   
  as a source of the geometry. In particular, K. Bronnikov \cite{Bronnikov1} presented \ssph\ \wh\ 
  solutions with or without an electric charge both in general relativity (GR) and a class of 
  scalar-tensor theories. Refs.\,\cite{kb95, Cle3, PriesleiGoulart, HyatHuang} describe wormholes with 
  electric or magnetic charges in the Einstein-Maxwell-dilaton theory. A family of dynamic solutions 
  of the Einstein-Maxwell-scalar theory that could describe an evolving charged black (white) hole 
  or wormhole has been obtained in \cite{JinboYang}. A magnetized rotating wormhole was obtained 
  in  \cite{Matos} as an exact solution to the Einstein equations. The recent papers 
  \cite{BlazquezSalcedo, kon21} have presented specific examples of static traversable wormholes in 
  Einstein-Dirac-Maxwell theory in four space-time dimensions, without explicit introduction of exotic
  matter, so that the Dirac fields themselves exhibit exotic properties \cite{our_comm}.
  Static, spherically symmetric wormhole solutions of asymptotically safe gravity have been 
  recently studied in \cite{AlencarNilton}.
  Wormholes have also been considered in GR with Chaplygin gas \cite{ChaplyginLobo}, 
  various models of phantom energy and quintessence, including models of matter with the 
  energy--momentum tensor (EMT) structure of a perfect fluid \cite{Sushkov1, Kuhfittig2, Lobo3, 
  Kuhfittig3, Sahoo, Kuhfittig4, kb17}. In \cite{kb17} it was shown that \ssph\ \whs\ with two flat or
  AdS asymptotic regions cannot exist in GR with any source matter with isotropic pressure,
  therefore, all such perfect-fluid \whs\ contain their exotic sources only in a bounded region, 
  surrounded by vacuum, with thin shells on the boundaries. In this connection, we should also 
  mention a large class of wormhole models built completely using the thin-shell formalism, where 
  the whole amount of exotic matter is concentrated on a thin shell; among them, Refs. 
  \cite{Visser2, Visser3} are probably the first examples.
   
  Much of the literature on \whs\ is devoted to studies of static, spherically symmetric configurations. 
  Their important generalization is connected with dynamic solutions. Dynamic wormholes have 
  been considered in various aspects, and one of the ways of building these models 
  \cite{Kuhfittig3, Kar, Kim, Roman, SushkovKim, SushkovZhang} is to add a time-dependent 
  scaling factor to an otherwise
  static metric. Dynamic models of wormholes were also constructed using the thin shell formalism
  \cite{Anzhong}. General properties of arbitrary dynamic wormholes are discussed in 
  \cite{Hayward2, Dynwh1}. 

  As mentioned above, a basic problem of \wh\ physics is the necessity of exotic matter for their
  existence, at least in the case of static configurations in GR \cite{MorTho, hoh-vis1} and a wide
  class of scalar-tensor and f(R) theories of gravity \cite{kb-star}. Rotational degrees of freedom
  are apparently able to replace exotic matter at \wh\ construction. At least, some examples of 
  rotating cylindrically symmetric \wh\ models in GR have been built without NEC violation  
  \cite{cyl1,cyl2,cyl3}; the recent examples with a static metric \cite{BlazquezSalcedo, kon21}
  contain a Dirac field involving spin; one can also mention exotic-free solutions in the Einstein-Cartan
  theory \cite{gal1,gal2} containing nonzero torsion.
  
  Dynamic \whs\ can also avoid NEC violation, at least in cases where a static early-time or late-time
  asymptotic behavior is absent. Such \wh\ models with cosmological-type metrics are known, 
  for example, in GR with \emag\ fields described by some special forms of nonlinear electrodynamics 
  \cite{Arellano, NEDwh}. 
  
  In this paper we study the possible existence of traversable wormholes in GR with 
  another classical and nonexotic form of matter, widely used in various problems of astrophysics 
  and cosmology, namely, dustlike matter, with or without an electromagnetic field. 
  The general spherically symmetric solution to the Einstein equations with dustlike matter was 
  obtained by Lema\^{\i}tre and Tolman in 1933-1934 \cite{tolman,Lemaitre33} and studied by 
  Bondi in 1947 \cite{Bondi47}; in the literature it is called the Tolman or LTB solution. Its 
  properties were further investigated, for example, in \cite{Christodoulou,Landau,Bambi}. 
  An attempt to obtain a wormhole using a special choice of arbitrary functions in this solution 
  was undertaken in \cite{IKS}. 

  Tolman's solution was extended to include electrically charged dust with a radial \emag\ field in 
  \cite{markov,vickers,bailyn,shik,khlest,lapch} (see also references therein), where the problem 
  was completely solved under some special conditions and considered on the level of integrals
  of the Einstein-Maxwell equations in the general case. The full set of solutions for arbitrary 
  charged distributions and arbitrary initial data was obtained by N.V. Pavlov \cite{pavlov} and
  studied in \cite{pav-kb}; further extensions to plane and hyperbolic symmetries were considered 
  in \cite{kb83a, kb83b, kb84}, see also references therein.
  
  The present study is restricted to configurations with an external magnetic (or electric) fields
  and electrically neutral dust, so that if we obtain a \wh, its every entrance will contain a 
  ``charge without charge'' \cite{Wheeler1, Wheeler2} due to a bundle of electric or magnetic lines 
  of force threading the throat. Similar models with radial magnetic fields and a particular choice 
  of initial data have been studied in \cite{Shatskii, hle-suh}, while here we consider the 
  conditions for \wh\ existence under general initial data.  

  It is worth noting that wormhole throats are defined in different ways by different authors. In the 
  static case, different definitions agree with each other whereas in dynamic models contradictions 
  may arise. General definitions of a throat including time-dependent metrics are considered, 
  for example, in \cite{Dynwh1, TomikawaIzumi, Bittencourt}. Following the papers 
  \cite{Kar, Kim, Roman}, we will use a definition that rests on the properties of the 3-geometry
  of spatial sections of a given space-time.
  
  It turns out that \wh\ throats in the solution under study can be located only in a T-region, i.e., 
  a nonstatic region like that inside a \bh. Therefore, solutions containing a throat can lead to 
  \whs\ as global entites only in a cosmological context, but if we try to inscribe such a solution 
  to an \asflat\ space-time, we can only obtain \bhs\ which also possess some properties of interest.
    
  The paper is organized as follows. In Section 2 we reproduce a derivation of the generalized Tolman 
  solution describing the evolution of a spherically symmetric cloud of dustlike matter in the presence 
  of an external electric or magnetic field, without an explicitly introduced source. In Section 3 
  we investigate the possible existence of throats and traversable wormholes.
  Section 4 describes some particular examples, and Section 5 is a conclusion.

\section{Tolman's solution with an electric or magnetic field}

  Let us consider a generalization of Tolman's solution \cite{tolman, Lemaitre33, Bondi47} 
  (to be called the $q$-Tolman solution), describing the evolution of a \sph\ cloud of 
  electrically neutral dustlike matter in the presence of an external electric or magnetic field.
  In what follows, for certainty, we prefer to speak of a magnetic field since it is more 
  realistic from an astrophysical viewpoint, though all solution to be discussed can be 
  interpreted as those with an electric field. 
 
  If we choose a comoving reference frame for neutral dust particles, it is also a geodesic 
  reference frame for them, and the metric can be taken in the synchronous form
\beq                                                    \label{ds-tol}
      ds^2 = d\tau^2 - \e^{2\lambda(R,\tau)} dR^2 - r^2(R,\tau) d\Omega^2,
\eeq
  where $\tau$ is the proper time along the particle trajectories labeled
  by different values of the radial coordinate $R$, $\lambda(R,\tau)$ and $r(R,\tau)$ are 
  functions of $\tau$ and $R$. 
		
  The energy-momentum tensor (EMT) of dustlike matter in the co-moving frame has the form 
  $T_{\mu}^{\nu [d]}=\rho u_{\mu} u^{\nu}$, 
  where $(u^{\nu})=(1, 0, 0, 0)$ is a velocity four-vector, and $\rho$ is the energy density. 
  The only nonzero component of the EMT of dust is $T^{0 [d]}_0 = \rho$, 
  while for the \emag\ field the EMT may be presented as
\beq   									                  \label{SET-e}
		{T\mN}^{[{\rm em}]} = \frac{q^2}{8\pi G r^4} \diag(1,\ 1,\ -1,\ -1),
\eeq  
  where $q$ may be interpreted as an electric or magnetic charge in proper units   
  \cite{Reissner,Nordstrom}. 
  
  The nontrivial components of the Einstein equations may be written in the form
\bearr
	   2r\ddot{r} + \dot{r}{}^2 +1 - e^{-2\lambda}r'{}^2 = \frac{q^2}{r^2},  \label{G11}
\yyy
		\frac{1}{r^2}(1 + \dot{r}^2 + 2 r\dot{r}\dot{\lambda})
       	- \frac{\e^{-2\lambda}}{r^2} (2rr'' + r'{}^2 - 2 rr'\lambda' )
            									= 8\pi G \rho + \frac{q^2}{r^4}, \label{G00}
\yyy
		      \dot{r}' - \dot{\lambda} r' = 0.			\label{G01}
\ear
  Equation (\ref{G01}) is easily integrated in $\tau$ giving
\beq 
		\e^{2\lambda} = \frac{r'{}^2}{1 + f(R)},                  \label{lam}
\eeq
  where $f(R)$ is an arbitrary function satisfying the condition
  $1 + f(R) > 0$. Substituting (\ref{lam}) into (\ref{G11}), we obtain
  the equation
\beq 							\label{ddr}
  			  2r \ddot{r} + \dot{r}^2 = f(R) + \frac{q^2}{r^2},
\eeq
  whose first integral is
\beq 									 		\label{dr}
			\dot{r}^2 = f(R) + \frac{F(R)}{r}- \frac{q^2}{r^2}.
\eeq
  This expression makes clear the physical meaning of the function $f(R)$:
  since $\dot{r}$ can be loosely understood as the radial velocity of a dust
  particle, $f(R)$ specifies the initial dust velocity distribution: for
  $f \geq 0$ it is the particle velocity squared at large $r$ ($f >0$ and
  $f=0$ correspond to hyperbolic and parabolic motion, respectively).
  If $f(R) <0$  (elliptic motion), the particle cannot reach infinity, and the 
  condition $\dot r=0$ in \rf{dr} shows the maximum value of $r$
  accessible to it.

  The meaning of the other arbitrary function, $F(R)$, becomes clear if we
  substitute (\ref{lam}) and (\ref{dr}) into the constraint equation (\ref{G00}). 
  We obtain
\beq
		\rho  = \frac{1}{8\pi G} \frac{F'(R)}{r^2 r'},       \label{rho}
\eeq
  or
\beq  												\label{F-tol}
		F(R) = 8\pi G \int \rho r^2 r' dR.
\eeq
  Assuming that in the initial configuration (before collapse) there was a
  regular center, so that $F =0$ at $r = 0$, we can write
\beq                                                              \label{M-tol}
      	F(R) = 2G M(R), \qquad M(R) = 4\pi \int_{0}^{r} \rho r^2 r' dR,
\eeq
  so that $M(R)$ is the mass function equal to the mass of a spherical body 
  including all matter inside the sphere of given
  radius $r$, and $F(R)$ is the corresponding Schwarzschild radius.
  If there is no regular center, such a direct interpretation becomes impossible.  
  However, the relation $F(R) = 2G M(R)$ can also be used in cases without
  a regular center (or even if there is no center at all). 
  
  Equation (\ref{dr}) can be further integrated, and the solution depends on the sign of $f(R)$:
\bearr    \nq                                         \label{tau+}
     f >0: \quad 
        \pm[\tau-\tau_0(R)] = \frac 1f \sqrt{fr^2 +Fr -q^2} 
         					-  \frac F{2f^{3/2}} \ln \Big(F + 2fr + 2\sqrt{f} \sqrt{fr^2 +Fr -q^2} \Big),
\yyy           \nq                                       \label{tau0}
     f = 0: \quad 
     		 \pm[\tau-\tau_0(R)] = \frac{2 \sqrt{Fr -q^2} (F r + 2 q^2)}{3 F^2},
\yyy              \nq                                 \label{tau-}
     f < 0: \quad 
     		 \pm[\tau-\tau_0(R)] =  \frac 1h \sqrt{-hr^2 +Fr -q^2} 
     		 			+ \frac {F}{2 h^{3/2}} \arcsin \frac {F-2hr}{\sqrt{F^2 - 4hq^2}}, 
     		 			\qquad h(R) := -f(R).
\ear
  Note that the elliptic model (\ref{tau-}) is admissible only in the case ${F^2-4hq^2\geqslant 0}$.
  The new arbitrary function $\tau_0(R)$ corresponds to a choice of clock
  synchronization between different spheres with fixed $R$ (Lagrangian spheres).

  To obtain a global solution including both a dust distribution and an external 
  electrovacuum region described by the \RN\ metric, it is necessary to describe such an
  electrovacuum region in a reference frame suitable for smooth matching to the 
  internal solution. This problem is easily solved since the external solution in a proper 
  form is obtained from the presently discussed solution (to be called ``q-Tolman solution'') 
  by putting $F(R) = \const$ whence $\rho =0$. The resulting solution reproduces 
  the \RN\ \cite{Reissner,Nordstrom} metric in a geodesic reference frame.
  A global configuration is then represented by the q-Tolman solution in which $F(R)$ 
  is variable in a certain range of $R$, say, $R < R_0$, and constant at larger $R$,
   so that $F(R) = F(R_0) = 2GM(R_0)$, $M(R_0)$ being the \Scw\ mass. 
  
  This makes possible an interpretation of the arbitrary function $M(R)$ independent 
  from any assumptions on a center of the configuration: at each $R$, it is the \Scw\ mass 
  of the external (electro)vacuum solution that can be joined to the internal one at this 
  particular value of $R$.

\section{Possible wormhole throats}
\def\th{_{\rm th}}

  As follows from \rf{rho}, to keep the density positive, we must require 
  $F'/r' > 0$, but it is not necessary to assume that both $F'$ and $r'$ are positive.
  This means that one cannot exclude the existence of regular minimum or maximum values
  of $r$ (at given $\tau$), which can be interpreted as throats and equators, respectively. 
  
  As already said in the Introduction, there are different definitions of a \wh\ throat suitable for  
  general time-dependent metrics. We will use the definition according to \cite{Kar, Kim, Roman}.
  So, let us call a {\bf throat} a regular minimum of the spherical radius $r(R, \tau)$ 
  at given $\tau$ (that is, in a fixed spatial section of our space-time), and let us try 
  to understand whether it is possible to build a \wh\ configuration based on the $q$-Tolman 
  solution. The first thing to do is to find out the conditions to be satisfied on a \wh\ throat.
  
  Considering a section $\tau = \const$, we can write the 3D spatial metric as
\beq                              \label{ds_3}
                   dl_{(3)}^2 = \frac{r'{}^2 dR^2}{1+ f(R)} + r^2(R) d\Omega^2.
\eeq   
 where $r(R)=r(R,\tau)|_{\tau=\const}$. 

  Using the freedom to choose any radial coordinate $R$, we might simply take 
  (at fixed $\tau$) $R = r$, but then, if $r$ has a minimum, it is not an admissible coordinate 
  precisely at this minimum, the throat. To avoid using such undesirable radial coordinates,
  we can take the manifestly admissible Gaussian coordinate $l$, such that $dl = |g_{RR}|^{1/2} dR$ 
  is the element of length in the radial direction. Then the throat conditions are 
\beq                              \label{th}
			\frac{dr}{dl} = 0, \qquad  \frac{d^2 r}{dl^2} > 0 
\eeq  
  (assuming a generic minimum and ignoring possible high-order ones, with a zero 
  second-order derivative). The first condition implies that at the throat, $R= R\th $,
\beq                              \label{th1}
			\frac{dr}{dl} = \sqrt{1+ f(R\th )} = 0 \ \ \then \ \ f(R\th ) = -1 
					\ \ \ {\rm or}\ \ \  h(R\th )=1.
\eeq      
  We immediately obtain that only elliptic models \rf{tau-} can be suitable for \wh\
  construction. Furthermore, since in general, to keep the metric \rf{ds-tol} 
  nondegenerate, we must have $1+f  = 1 -h >0$, and $h = 1$ may occur only at a 
  particular value of $R$, it is clear that $R=R\th $ must be a maximum of $h(R)$, 
  and it should be $h'(R\th ) =0$ and $h''(R\th ) <0$. The second condition \rf{th} leads to 
\beq                             \label{th2}
			\frac{d^2 r}{dl^2}\Big|_{R=R\th } = -\frac{h'}{2r'}\Big|_{R=R\th } > 0.
\eeq   
  Thus $r'$ must vanish at $R=R\th $ along with $h'$.
  A finite limit of $h'/r'$ may be found using the L'Hospital rule. The conditions (\ref{th1}) 
  and (\ref{th2}) impose restrictions on the functions $F(R)$ and $h(R)$. 
  
  A special case of functions $F(R)$ and $f(R)$ satisfying the throat conditions (\ref{th1}) 
  and (\ref{th2}) was considered in \cite{IKS}. In this paper, we consider the general case
  without specifying the functions $F(R)$ and $f(R)$ and test the traversability of the models. 
  The point is that the existence of a throat is only a necessary condition for having a wormhole:
  it is also necessary to require the absence of horizons which are only one-way traversable,
  and even without horizons the traversability may be destroyed by the time evolution. 

  For the elliptic models \rf{tau-} it is convenient to introduce the parametric representation,
\bear                 			 \label{eta}
			r \eql \frac {F}{2h} (1 - \Delta \cos \eta), 
\nn
			\pm[\tau-\tau_0] \eql \frac {F}{2h^{3/2}}(\eta - \Delta \sin\eta),
						\qquad  \Delta = \sqrt{1 - \frac{4 hq^2}{F^2} },
\ear  
  where $0 < \Delta \leq 1$, and $\Delta =1$ corresponds to Tolman's solution without 
  an electromagnetic field. We see that at $q \ne 0$, hence $\Delta < 1$, the model 
  is free from singularities characterized by $r=0$. On the other hand, in the special case
  $\Delta =1$ ($q=0$), the solution reduces to Friedmann's closed isotropic model filled 
  with dust \cite{Landau} under the assumptions
\beq       \label{Fri1}
	    F(\chi) = 2 a_0 \sin^3\chi, \qquad   h (\chi) = \sin^2 \chi, \qquad a_0 = \const
\eeq  
  (the radial coordinate $R = \chi$ is here a ``radial angle'' of a 3D sphere), and we have
\beq         \label{Fri2}  
  		r = r(\eta,\chi) = a(\eta) \sin \chi,  \qquad a(\eta) = a_0 (1-\cos \eta),
\eeq  
  $a(\eta)$ being the cosmological scale factor. 
    
  An important question for our \wh\ study is the allocation of R- and T-regions
  separated by apparent horizons \cite{HawkingEllis}. The answer follows from the definition of 
  R- and T-regions, according to which a T-region is the one where the gradient of 
  $r(R, \tau)$ is timelike, an R-region where it is spacelike, and it is null at a horizon. Evidently,
\beq                                                       \label{grad}
	   r^{,\alpha}r_{,\alpha} = \dot{r}^2 - \e^{-2\lambda}r'{}^2
			= -1 + \frac {F(R)}{r} - \frac {q^2}{r^2},
\eeq
 where the second equality sign follows from \rf{lam} and \rf{dr}. Recalling that 
 $F(R) = 2GM(r)$, we observe that the R- and T-regions occur in full similarity with
 the \RN\ space-time, where, instead of the mass of a central body, there stands 
 the current value of the mass function $M(R)$.  
   
 Recalling that a possible \wh\ throat can only occur at $h = 1$ and substituting  the 
 condition \rf{th1} and
 $r(R, \tau)$ from \rf{eta}, we finally obtain that at such a throat
\beq                                                       \label{grad-th}
	r^{,\alpha}r_{,\alpha} \Big|_{R\th}
			= \frac 1{r^2} \sin^2 \eta \Big(\frac{F^2}{4} - q^2 \Big) \geq 0.
\eeq  
  It means that a seeming \wh\ throat is located, in general, in a T-region, and only at 
  $\eta = \pi n,\ n\in \N$ (at which the radius $r$ for given $R$ takes extremal values 
  as a function of $\tau$), the throat coincides with a horizon. 
  
  This means that a would-be \asflat\ \wh\ configuration that could be constructed from
  the $q$-Tolman solution cannot be traversable and, if matched with an external \RN\ 
  solution at some value of $R$, actually represents a \bh. 
  This also applies to the special case $q=0$, the absence of a magnetic field, such that 
  we return to the Tolman solution: as follows from (\ref{grad-th}), this solution
  also does not admit traversable wormholes with flat asymptotic regions. 
  If it is matched to an external Schwarzschild solution at some value of $R$, we obtain
  a black hole model with a dust source. In particular, this applies to the seemingly wormhole
  model considered in \cite{IKS}.
  
  It may seem that such configurations with throats can be used to construct regular
  \bh\ configurations. We will see, however, that even though $r >0$ for all Lagrangian 
  spheres if $q\ne 0$, all solutions with a throat exist in a finite time interval, beginning and 
  ending with shell-sticking singularities due to $r'=0$. Another point of interest can be 
  the existence of dynamic cosmological \whs, which can be obtained if we do not match the 
  q-Tolman solution to a \RN\ one on both sides of the throat but, instead, require  
  Freidmann-Robertson-Walker asymptotics on both sides, see Section 4.
	
  One more observation is that, as already follows from \eq \rf{dr}, the elliptic mode of dust 
  motion is only compatible with charges $q < F/2 = GM$, corresponding
  to the \bh\ branch of the \RN\ solution. Larger charges, which could seem to favor 
  the existence of \whs, require the hyperbolic mode, which is in turn in conflict with 
  the condition  \rf{th1}. 
  
  As follows from the expression \rf{rho} for the dust density, $\rho \to \infty$ if either $r\to 0$
  or $r' \to 0$, except for the case where both $r'\to 0$ and $F' \to 0$ at finite $r$, such that 
  $F'/r'$ remains finite, precisely the situation of interest for us at a \wh\ throat. The absence of 
  a singularity under these conditions is confirmed by the expression for    
  the Kretschman scalar $\cK$\footnote
  		{In some books we can find a value of $\cK$ four times larger, for example \cite{Bambi}. 
		 However, one can verify that \eqn{Kre} gives the correct result for the Friedmann solution
		 \rf{Fri1}, \rf{Fri2}: $\cK=60 a_0^2/a^6$.}
\beq                       \label{Kre}
		\cK(R,t) =3\frac{{F'}^2}{{r'}^2r^4}-8\frac{F'F}{r'r^5}+12\frac{F^2}{r^6}
					+20\frac{F'q^2}{r'r^6} - 48\frac{Fq^2}{r^7}+56\frac{q^4}{r^8}.
\eeq

  We have seen that possible throats require simultaneous vanishing of the three
  derivatives $r'$, $F'$ and $h'$, with finite limits of the ratios $F'/r'$ (see \rf{rho} and \rf{Kre}) 
  and $h'/r'$ (see \rf{th2}), and we also try to keep the density $\rho$ positive.
  
  By \rf{eta}, the derivative $r'$ on a constant-$\tau$ spatial section of our space-time is given by
\bearr 
		r'=\frac{Fh'N_1(R,\eta) + 2h F' N_2(R,\eta)} {4\Delta h^2(1-\Delta\cos\eta)}, 
\nnnv
		N_1(R,\eta) = \cos\eta -3\Delta + 3\Delta^2 (\eta\sin\eta+\cos\eta) 
							 + \Delta^3 ( -2 + \cos^2\eta),
\nnnv					
		N_2(R,\eta) = 	- \cos\eta + 2\Delta- \Delta^2(\cos\eta + \eta\sin\eta).\label{r_R}
\ear
  Since at a throat $R=R\th $ the quantities $F'/r'$ and $h'/r'$ are finite and nonzero 
  (though have different signs), $r'$, $h'$ and $F'$ are infinitesimal quantities of the same order 
  of magnitude near such a throat.
  
  We can summarize that on a throat $R=R\th $ we must have
\bearr                    \label{throat}
		h = 1, \quad\ h'=0, \quad\ h'' < 0,
\nnn		
		F' = 0, \quad\ r' =0,\quad\   \frac{h'}{r'} < 0, \quad\  \frac{F'}{r'} > 0. 
\ear
   Also, we have everywhere $F^2 - 4hq^2 > 0$ and $\Delta \leq 1$.

\begin{figure}
 \centering   
 \includegraphics[scale=0.7]{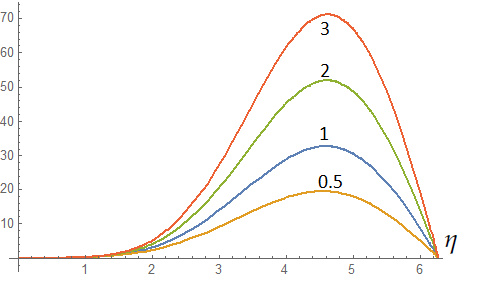}
 \includegraphics[scale=0.7]{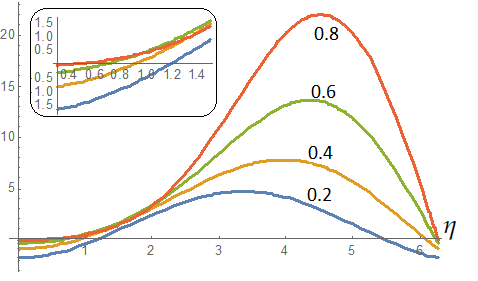}
\caption{\small
	Time dependence of $N_* = 2N_2-BN_1$. Left: $\Delta = 1,\ B=0.5,1,2,3$;  
	right: $B =1,\ \Delta=0.2, 0.4, 0.6, 0.8$; for other values $B$ and $\Delta$ the plots 
	looks similarly. The inset on the right panel clarifies the points where $N_* =0$.
	Note that $\rho$ and $d^2 r/dl^2$ at $R=R\th $ are positive or negative simultaneously 
	with $N_*$. \label{N*}}
\end{figure}

  For a further analysis, consider the limit $\lim\limits_{R\to R\th }\dfrac{Fh'}{F'h}=-B$ 
  with $B = \const  \geq 0$. Then we have on the throat 
\bearr 
		r'\Big|_{R\th }  = \frac{F'(2N_2 - B N_1)} {4\Delta (1-\Delta\cos\eta)},
\ear
 and it vanishes only where $F'=0$ or $N_* = 2N_2 - BN_1 = 0$. The density (\ref{rho}) and 
 the second-order derivative $d^2r/dl^2$ (\ref{th2}) have there the values 
\bearr          \label{rho_th}
		\rho(R\th ,\eta) =  \frac {\Delta (1 - \Delta\cos\eta)} {2\pi G r^2 (2N_2-BN_1)}\bigg|_{R\th },
\\ \lal                    \label{d2r_th}
		\frac{d^2 r}{dl^2}\bigg|_{R\th } =  \frac{2 B}{F} 
					\frac{\Delta( 1-\Delta\,\cos \eta)}{2N_2-BN_1}\bigg|_{R\th }.
\ear
  These quantities blow up when $N_* = 2N_2 - BN_1 = 0$, while all other factors 
  are strictly positive ($F > 0$ by assumption). However, the quantity $N_*$ as a function 
  of $\eta$ has, in general, an alternating sign, as follows from the expressions 
  (see also Fig.\,\ref{N*}). 
\bearr
	N_*\Big|_{\eta=0,2\pi} = - (1-\Delta)^2 [2+B (1-\Delta)] \leq 0, 
\nnn	
	N_*\Big|_{\eta=\pi} = (1+\Delta)^2 [2+B (1+\Delta)] > 0.
\ear
  Therefore we have inevitably $N_* = 0$, hence a singularity, at (at least) two values of $\eta$, 
  say, $\eta_1$ and $\eta_2$, in the range $\eta \in (0, 2\pi)$ for any $\Delta < 1$ ($q \ne 0$). 
  In the case $q=0$, $\Delta=1$ (pure dust), $N_*$ vanishes at $\eta_{1,2} =0,\,2\pi$ and is 
  positive at $\eta \in (0, 2\pi)$. The quantity $N_*$ is plotted in Fig.\,\ref{N*} for some values
  of $B$ and fixed values of $\Delta$. Plots of the density $\rho$ will be 
  shown below for a specific example. 

  We conclude that a nonsingular evolution of a throat $R=R\th $ with positive density $\rho$ 
  can take place at times $\eta_1 < \eta < \eta_2$ at which $N_* > 0$. For other Lagrangian 
  spheres $R=\const$ similar time limits will be different due to $R$ dependence of $F$ and $h$. 
  
\section{Some particular models}
\subsection{A special \wh\ solution}

  Let us try to construct an example of a q-Tolman \wh\ solution, choosing the following 
  simple functions of $R$ in agreement with the requirements \rf{throat}:
\beq                												 \label{ex1}
		h(R)=\frac{1}{1+R^2}, \quad \ f(R)=2b (1+R^2) \ \ \then \ \ 
		\Delta=\sqrt{1-\frac{q^2}{b^2(1+R^2)^3}}, 
\eeq
   ($b =\const > 0$, specifying a length scale), so that 
\beq                												\label{ex1-r}
		r(R,\eta) = b(1+R^2)^2 (1 - \Delta \cos \eta),
				\qquad
		r'(R,\eta) = \frac{bR (1+R^2) (2N_2 - N_1)}{\Delta(1-\Delta \cos \eta)}, 
\eeq
  with $N_{1,2}$ defined in \eqn{r_R}. The throat is located at $R=0$, and the whole
  solution is symmetric with respect to it. The density $\rho$ (\ref{rho}) and the quantity 
  $d^2r/dl^2$ at $R=0$ then read
\bear  			\label{rho_th_ex}
	\rho(R,\eta) \eql \frac{\Delta}{2\pi G b^2 (1+R^2)^5 (1-\Delta\cos\eta) (2N_2-N_1)},
\\                        \label{d2r_th_ex}
	\frac{d^2r}{dl^2}\bigg|_{R=0} \eql \frac{\Delta\,( 1-\Delta\,\cos \eta)}{b(2N_2-N_1)}\bigg|_{R=0}.
\ear
  As already noted, different signs of the derivatives of $h(R)$ and $f(R)$, under the condition
  $2N_2(R,\eta) - N_1(R,\eta) > 0$, provide the validity of the throat conditions \rf{throat} at $R=0$
  and, by continuity, in some its neighborhood, but the same is not guaranteed at all $R$ and $\eta$. 
  
\begin{figure}[h]
\centering
\includegraphics[scale=0.35]{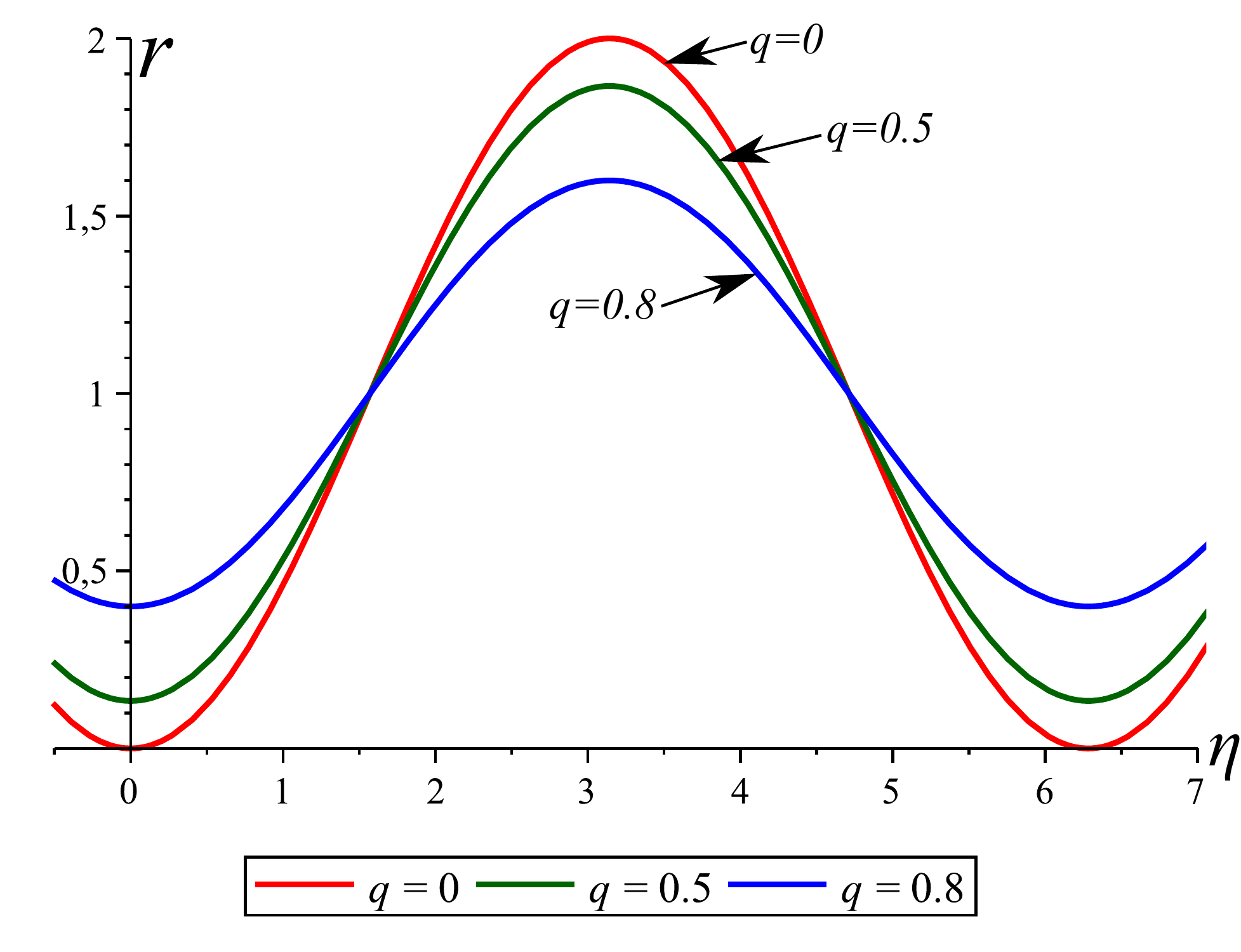}\ 
\includegraphics[scale=0.35]{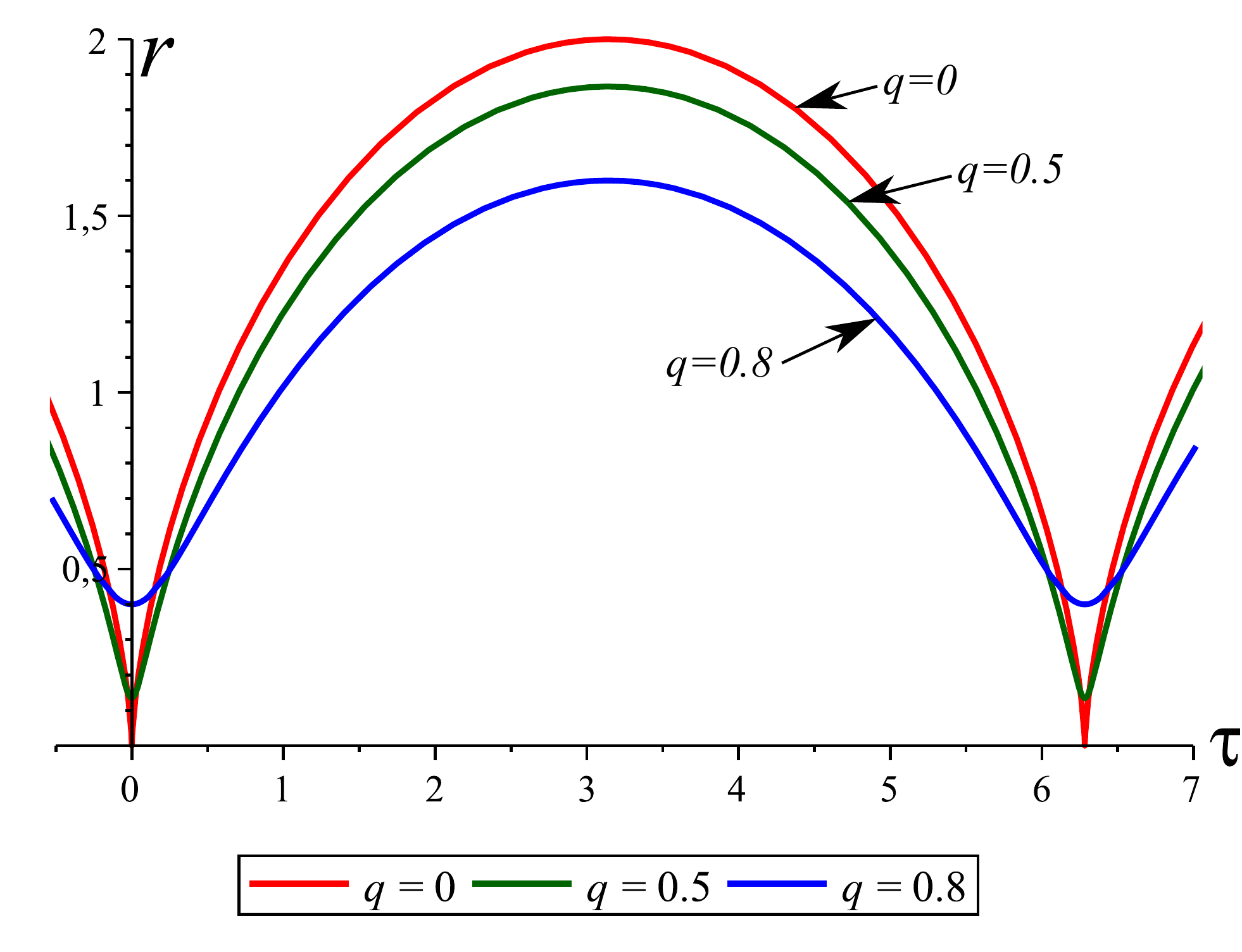}
\caption{\small
	Time dependence of the throat radius for $q=0, 0.5, 0.8$ in terms of $\eta$ (left) 
	and in terms of $\tau$ (right). \label{fig2}}
\end{figure}

  The time dependence of the throat radius is shown in Fig.\,\ref{fig2}, and the density $\rho$ on
  the throat in Fig.\,\ref{rho,r_throat} for different values of $q$, where dashed lines show 
  the asymptotes of the function.
  Finite positive density values are observed for a limited period of time $\eta \in (\eta_1, \eta_2)$ 
  while $2N_2-N_1 >0$, between two singularities where $\rho$ and $\cK$ diverge. Outside this 
  interval, in the case $q\neq 0$, the density changes its sign along with $d^2r/dl^2$, therefore 
  the throat conditions hold together with the condition $\rho > 0$. 

  Thus we observe a good \wh\ behavior of our solution at the time interval $\eta \in (\eta_1, \eta_2)$. 
  With decreasing charge, this interval increases; and at $q=0$ we have $\eta_1=0$, $\eta_2=2\pi$. 
  Outside the throat (at $R\neq 0$), the plots look similarly, but the singularities occur at other 
  time instants.

\begin{figure}[h]
\centering
\includegraphics[scale=0.5]{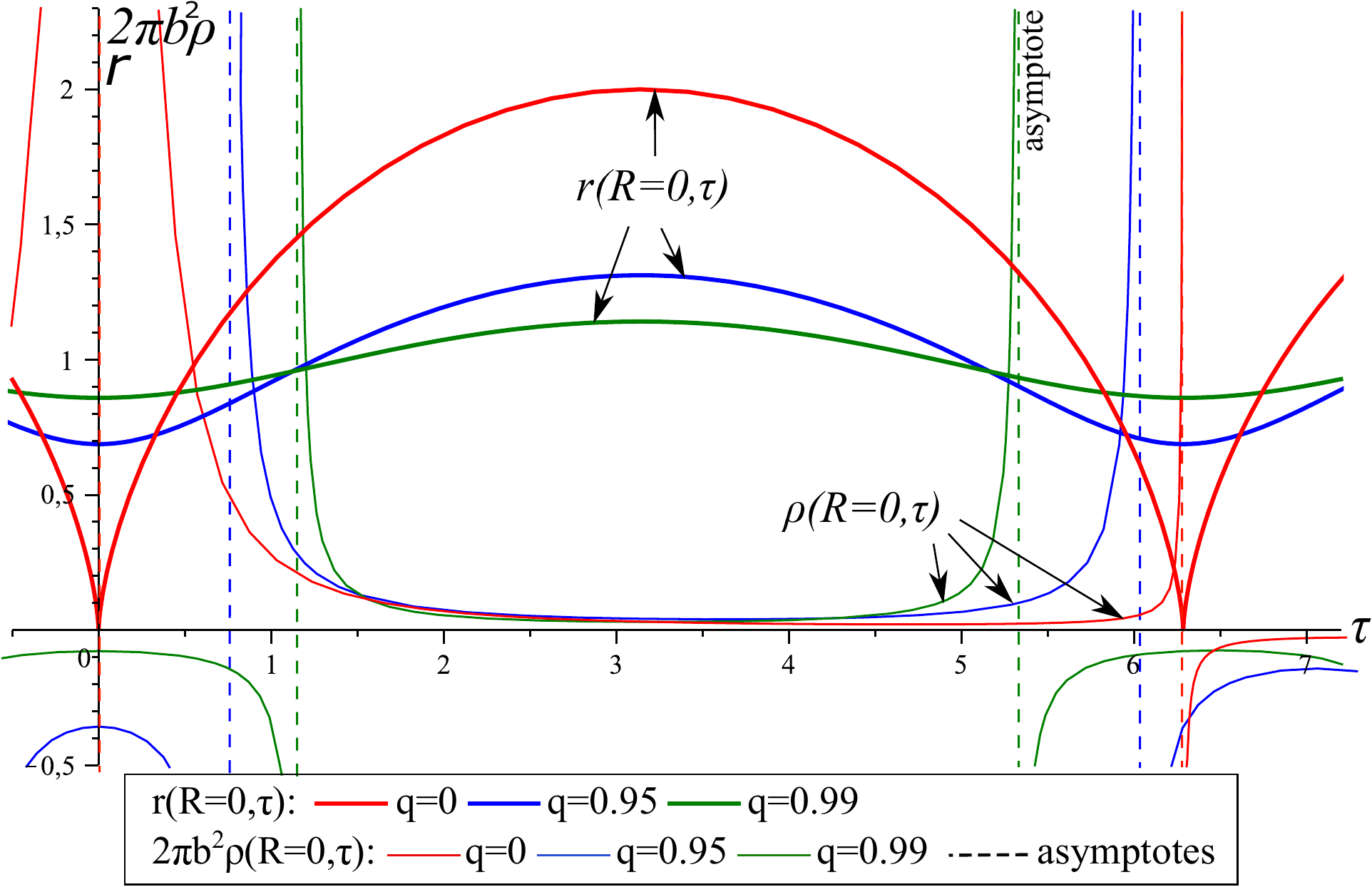}
\caption{\small
	Time dependence of the functions $2\pi b^2\rho$ (thin lines) and $r(0,\tau)$ (thick lines) 
	on the throat $R=0$ for $q=0,\,0.95,\,0.99$ in the model \rf{ex1}. 
	Dashed lines show the asymptotes. For other values of $q$ the plots look in a similar way. 
	\label{rho,r_throat}}
\end{figure}   

\subsection{Wormholes in a dust-filled universe}

   Now, let us look how such a \wh\ solution can be inscribed into the closed isotropic 
   cosmological model characterized by the relations \rf{Fri1}, \rf{Fri2}. To that end we 
   note that to match two q-Tolman space-times, characterized by different functions 
   $F(R)$ and $h(R)$, at some value of the radial coordinate $R = R^*$, one should above 
   all identify the hypersurface $\Sigma$ ($R = R^*$) as it is viewed from different sides 
   since in unified space-time it should be a single hypersurface. This means that the metric
   tensor should be continuous on $\Sigma$. With the metric \rf{ds-tol} it means simply 
   $[r^2(R,\tau)] =0$ (where the square brackets, as usual, denote jumps at a transition
   across $\Sigma$), while by \rf{ds-tol} $g_{\tau\tau} \equiv 1$ on both sides.
   
   Next, according to \cite{darmois, israel}, discontinuities in the second quadratic form 
   of $\Sigma$ correspond to a nonzero stress-energy tensor of the surface matter content, 
   and to avoid its presence we should require that it should have no jumps.
   For the metric \rf{ds-tol}, this requirement leads to $[\e^{-\lambda}g'_{\tau\tau}] =0$
   (which trivially holds) and $[\e^{-\lambda} r'] =0$. Taking into account \rf{lam} and \rf{eta}, 
   we obtain
\beq                          \label{junc}
           [r] = 0, \quad\  [\e^{-\lambda} r'] =0 \ \ \then \ \ [h] =0, \quad [F]=0, \quad [q]=0.
\eeq    
  Thus to match two q-Tolman solutions on a surface $\Sigma$ ($R=R^*$), we must 
  identify the values of $F(R^*)$ and $h(R^*)$ as well as the parameter $q$ in these solutions. 
  Then, by virtue of \rf{eta}, these matching conditions are valid at all times at which both 
  solutions remain regular. Also, there is no necessity to adjust the choice of the radial
  coordinates on different sides of $\Sigma$ because both quantities $r$ and $\e^{-\lambda} r'$ 
  are insensitive to reparametrizations of $R$, and the same is true for the functions $h(R)$ 
  and $F(R)$ which behave as scalars at such reparametrizations.
  
  Now, let us apply the conditions \rf{junc} to the solution \rf{Fri1}, \rf{Fri2} with $q=0$ 
  and \rf{ex1}, \rf{ex1-r} on the other, also putting $q=0$, and specifying the junction surface
  by some values $\chi = \chi^*$ and $R = R^*$. We then obtain
\beq                    \label{junc1}
			R^* = \cot \chi^*, \qquad b = a_0 \sin^5 \chi^*.
\eeq   
  that provides matching at $R^* > 0$. Since the functions involved in \rf{ex1} are even, 
  the same kind of matching can be implemented at $R^* < 0$. The whole composite model
  thus consists of two evolving closed Friedmann universes filled with dust and connected 
  through a \wh, forming a dumbbell-like configuration.
  
  Some numerical estimates are in order. Assuming $a_0 \sim 10^{28}$ cm, approximately the 
  size of the visible part of the Universe, let us require, on the other hand, that the throat 
  radius $r_{\rm th} \sim b$ should substantially exceed the Planck length, $b \gg 10^{-33}$ cm. 
  Then, using \rf{junc1}, we easily obtain 
\beq   
			\sin \chi^* \gg 10^{-12}, \qquad   R^* \ll 10^{12}, 	
\eeq
  which implies $r^* = r(R^*) \gg 10^{16}$ cm $\sim 0.01$ light year. It means that the wormhole 
  region should be rather large on the astronomic scale. If we assume $r^* \sim 100$ light years, 
  a stellar cluster size, it follows $\chi \sim 10^{-8}$ and $r_{\rm th} \sim 10^{20} l_{\rm pl}
  \sim 10^{-13}$\,cm, the throat having the size of an atomic nucleus.  If the \wh\ region is 
  as large as a galaxy, $r^* \sim 30\ {\rm kpc} \sim 10^{23}$\,cm, we obtain $\chi \sim 10^{-5}$,
  resulting in $r^* \sim 10^3$ cm, a throat of macroscopic size.
  
  The density \eqn{rho} for our model \rf{ex1} with $q=0$ is given by 
\beq                       \label{rho1}
  		\rho = \frac{1}{2\pi G b^2}\, \frac{1}{(1+R^2)^5 (1-\cos\eta)(2N_2 - N_1)}
\eeq   
  where $2N_2 - N_1$ is, for $\Delta =1$, a function of $\eta$, positive in the whole
  range $\eta \in (0, 2\pi)$. The time-dependent factor in \rf{rho1} can be approximated 
  by $1/10$, which is really true in a larger part of the time interval, and then,
  by order of magnitude, we have
\beq
             \rho \sim \frac {10^{-30}} {(1+R^2)^5 \sin^{10}\chi^*} \ \rm \frac{g}{cm^3}. 
\eeq  
   On the throat $R=0$ this gives large values: about $10^{50}\ \rm g/cm^3$ if
   $\chi^* = 10^{-8}$ and about $10^{20}\ \rm g/cm^3$ if $\chi^* = 10^{-5}$, much more
   than the nuclear density. On the other hand, on the junction surface $R =R^*$, since 
   $R^* \sim 1/\chi^*$, the density turns out to have a  universal value 
   independently from $\chi^*$,  $\rho \sim 10^{-30}  \ \rm g/cm^3$, of the order of 
   mean cosmological density.

\subsection{A wormhole-like structure inside a \Scw\ or \RN\ black hole}

  Let us now discuss the dynamics of our model with a throat assuming that dust matter 
  is distributed in some spherical layer containing a throat, and $R =\pm R^*$ is the radial 
  coordinate of the outer layer (Lagrangian sphere) on which the q-Tolman solution is matched 
  to the external \RN\ solution (or \Scw\ in the case $q=0$).
  Thus the external metric is
\beq                         \label{RNm}
		ds^2 = A(r) dt^2-\frac{dr^2}{A(r)} - r^2 d\Omega^2,
\qquad
		 A(r)=1-\frac{2M}{r}+\frac{Q^2}{r^2}, 
\eeq  
  where $M$ and $Q$ are the mass and charge of the source (using units in which $G=1$), 
  and we should consider the case $M > |Q|$ when the \RN\ space-time contains an event horizon 
  with $r = r_+ = M + \sqrt{M^2-Q^2}$ and a Cauchy horizons with $r = r_- =M - \sqrt{M^2-Q^2}$. 
  Naturally, there must be such an outer surface on each side from the throat, with 
  maybe different masses $M$ in the external regions but with the same charge $Q$. 
  We can also construct a model with a closed universe, as described above, on one side 
  from the throat and empty \asflat\ space on the other, but here we do not consider such models.  
  
  As already discussed in Section 2, the \RN\ solution is obtained from the q-Tolman solution
  as its special case such that $F(R) = 2M = \const$ and $q = Q$. According to \rf{junc},
  we must also have the jump $[h(R)] =0$ on the junction surfaces. Apart from this junction 
  condition, the function $h(R)$ remains arbitrary but can be given a desired form by accordingly 
  choosing the radial coordinate $R$, and with any such choice we are remaining in the 
  geodesic reference frame in \RN\ space-time. 

  Consider the motion of the outer surface of this thick dust layer from the initial singularity
  to the final one. As follows from the examples we have considered, both singularities 
  for any Lagrangian surface in models under consideration (that is, containing throats)
  are located in a T-region, where the gradient of $r(R, \tau)$ is timelike. It means, in particular,
  that dust particles belonging to an outer surface are initially located inside a white hole and 
  move in the direction of growing $r$. Further on they reach a maximum of $r$, and it is
  easy to prove that such a maximum occurs in the R-region, hence outside the event horizon 
  $r = r_+$. Indeed, recall the integral \rf{dr} of the equations of motion and assume 
  there the extremum condition $\dot r =0$, to obtain for the surface $R=R_s$
\beq                      \label{r_max}
		h(R) = \frac{2M(R)}{r} - \frac{q^2}{r^2} = 1 - A(r) \then  A(r_{\max}) = 1 - h(R)
\eeq   
  because the surface belongs to both internal and external regions. Since $h (R) < 1$ 
  for all $R$ except the throat, we obtain $A(r_{\max}) >0$, that is, an R-region.  
  
  After reaching $r_{\max}$, the surface layer shrinks again towards the event horizon 
  $r = r_+$, thus entering a \bh. The whole process takes a finite proper time $\tau$ of the
  surface particles, but from the viewpoint of a distant observer, the ``star'' of dust emerges 
  from the horizon in the infinitely remote past and collapses back in the infinitely remote future.

  All that happens in the same way irrespective of whether or not there is a \wh-like structure 
  inside the horizon. If this structure is present, it means that there is another external \RN\
  or \Scw\ space-time on the other side of the throat. One more specific feature of models 
  with a throat is that with $q \ne 0$ the whole evolution of all Lagrangian spheres 
  lasts a finite proper time interval between two singularities at which $r' \to 0$, the 
  so-called shell-crossing, or shell-sticking singularities, while in the general case of 
  q-Tolman solution, completely regular oscillating configurations are possible (recall that 
  with $q\ne 0$ the \emag\ field prevents reaching a singularity $r =0$ at which a 
  Lagrangian sphere shrinks to a point, see, e.g., Fig.\,\ref{rho,r_throat} that illustrates the 
  model evolution for a particular choice of the parameters in \eqn{ex1}.
  
  An important feature of models with a throat is that the Cauchy horizon $r = r_-$ is 
  never reached by the outer surface since the throat is a minimum of $r$ at each $\tau$
  and is located in a T-region between $r_-$ and $r_+$.
  
  A question of interest concerns the evolution of a throat $R = R\th$. We know that the
  throat is located in a T-region while a maximum of $r(R, \tau)$ is reached in an R-region
  for all $R \ne R\th$. How can it happen that the throat radius $r\th = r(R\th, \tau)$ has a 
  maximum? The answer is that a maximum of $r\th$ is reached precisely at the horizon, 
  a boundary between the R- and T-regions, and it is possible because $h(R\th) =1$,
  so the reasoning used about the surface layer (see \eqn{r_max}) does not work.
  In terms of the parametrization \rf{eta}, we have a maximum of $r\th$ at $\eta = \pi$, 
  when $r^{,\alpha}r_{,\alpha} =0$.
  
\subsection{Photon motion across the dust layer}

  Now we would like to consider the radial motion of light in the model (\ref{ex1}) of a dust 
  layer, assuming that it is bounded by $|R| < R^*$ and is located between two copies of 
  a \RN\ or \Scw\ (if $q=0$) space-time. It is clear that a photon radially falling to such a 
  \bh\ reaches the throat and has no other way than to travel further in the direction of another 
  universe or maybe a distant part of the same universe. The question is whether or not it will go 
  out from the dust layer in this ``other'' universe before the whole configuration reaches its final 
  singularity. In other words, is the \wh\ (or the \wh\ part of space-time) traversable.
  
  The radial null geodesic equation has the form following from the condition $ds =0$ 
  applied to the metric \rf{ds-tol},
\beq
	\frac{dR}{d\tau}=\pm\frac{\sqrt{1-h}}{r'}=\pm \frac{(\sign R)
	\Delta(1-\Delta\cos\eta)}{(1+R^2)^{3/2}(2N_2-N_1)},			\label{light_q}
\eeq
  where we have assumed $b=1$, and the minus and plus signs correspond to ingoing and
  outgoing light rays, respectively, in the region $R>0$. Of interest is the time interval 
  where $2N_2-N_1>0$, in which the right-hand side of \rf{light_q} is finite and nonzero, 
  therefore, in particular, the photon reaches the throat at finite proper time. By symmetry
  of the model ($R\leftrightarrow -R$), it is sufficient to consider the motion from positive to  
  negative values of $R$, choosing the minus sign in the region $R>0$ (the ingoing direction) 
  and the plus sign for $R<0$, which thus cancels the function $\sign R$. Thus we get 
\beq  			\label{R_tau}
	\frac{dR}{d\tau}= - \frac{\Delta(1-\Delta\cos\eta)}{(1+R^2)^{3/2}(2N_2-N_1)}.
\eeq

\begin{figure}[t]
\centering
\includegraphics[scale=0.7]{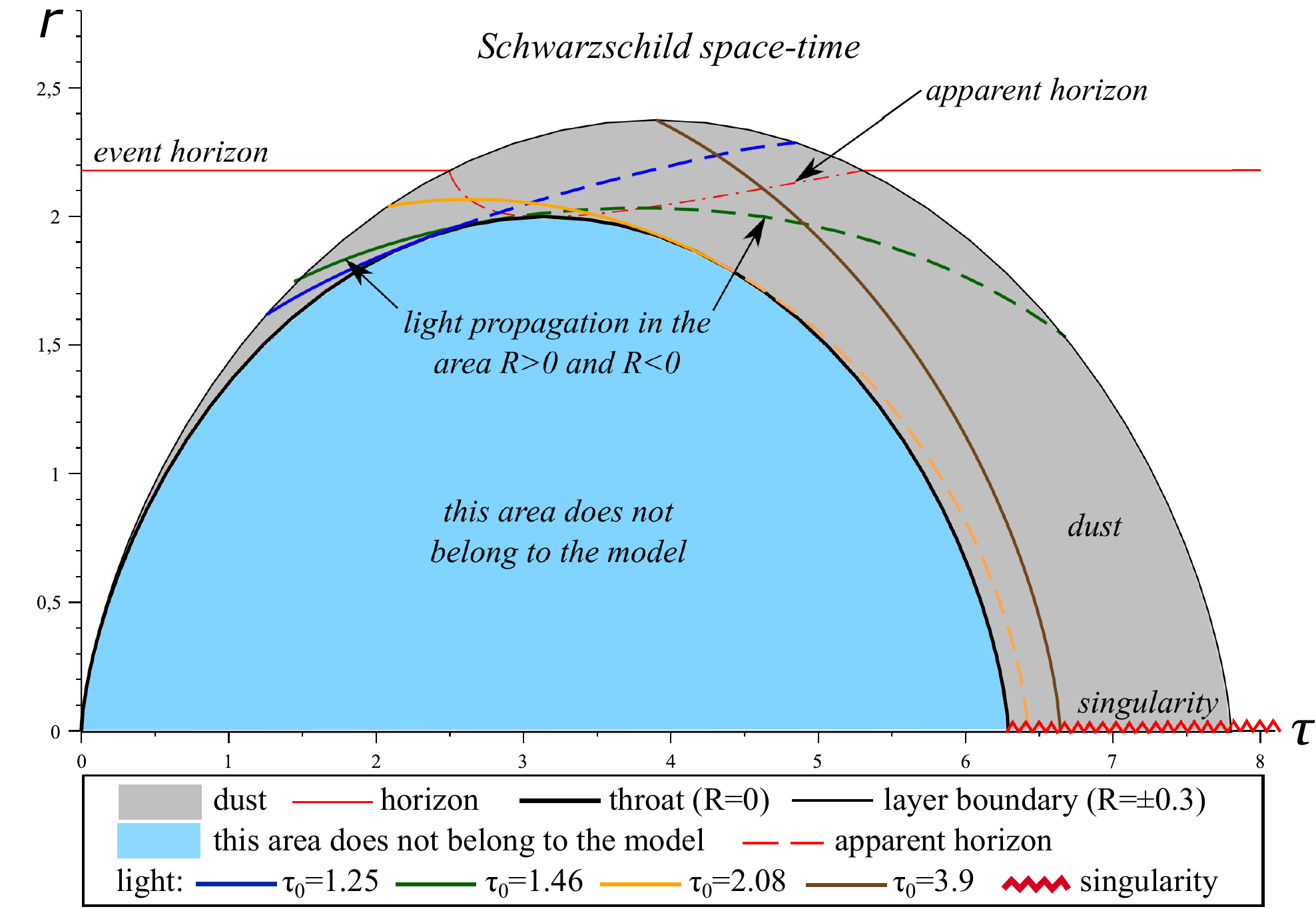}
\caption{\small
	The figure shows: 
	(1) The time dependence $r(\tau,R)$ for shells (Lagrangian spheres) $R=\const$ in the dust layer (gray area), 
	the throat $R=0$ and the outer surface $R=\pm 0.3$ in the case $q=0$. 
	(2) The time dependence $r(\tau)$ of ingoing radial photons coming from the outer shell $R=0.3$ towards the 
	throat $R=0$ at different times $\tau_0$; reaching the throat, a photon path passes from the region 
	$R>0$ (solid line)	to the region $R<0$ (dashed line);
     (3) The singularity, presented by a broken red line. 
     (4) The event horizon in vacuum and the apparent horizon inside the dust layer, presented by 
     solid and dashed red lines, respectively. \label{light_q0}}
\end{figure}
  Let us calculate the time derivative of the spherical radius $r(\tau,R(\tau))$ along the light 
  ray $R=R(\tau)$:
\beq                          \label{r_tau}
	\frac{d r(\tau,R(\tau))}{d\tau}  = \frac{\d r}{\d \tau}+\frac{\d r}{\d R}\frac{dR}{d\tau}
		= \pm\sqrt{-h+\frac{F}{r}-\frac{q^2}{r^2}}\pm \sqrt{1-h},
\eeq
  where $R=R(\tau)$ describes the radial motion of a photon according to (\ref{light_q}), 
  the first $\pm$ sign corresponds to expansion (plus, at $\eta\in(0,\pi)$) or contraction 
  (minus, at $\eta\in(\pi,2\pi)$) of the dust matter, while the second $\pm$ sign corresponds 
  to an ingoing (minus) or outgoing (plus) photon, that is, growing or falling values of $R$ along the path. 
  Note that when a photon reaches the throat ($h(R\th)=1$), it moves tangentially to the paths 
  of dust particles: $\d r(\tau,R(\tau))/\d\tau \equiv {\d r}/{\d \tau}$. 
  
  Curiously, it can happen that the quantity $r(R, \tau)$ for a radially moving photon,
  as a function of time, can have an extremum even though $\d R/\d \tau$ has a constant sign
  (a photon crosses the Lagrangian spheres in a definite direction).
  From \eqn{r_tau} it follows that such an extremum of $r$, at which $d r(\tau,R(\tau))/d\tau=0$,
  can take place only if the $\pm$ signs before the two terms are opposite, 
  which means that an extremum of $r$ is only possible at the contraction stage ($\eta > \pi$) for 
  an outgoing photon or at the expansion stage ($\eta < \pi$) for an ingoing photon. Specifically, 
  we have $d r(\tau,R(\tau))/d\tau=0$ at a surface where 
\beq
			r^2 - Fr + q^2 = 0, 
\eeq  
  that is, at the apparent horizon inside the dust layer, a surface where $r^{,\alpha} r_{,\alpha} =0$ in the metric 
  \rf{ds-tol}.

\begin{figure}
\centering
\includegraphics[scale=0.7]{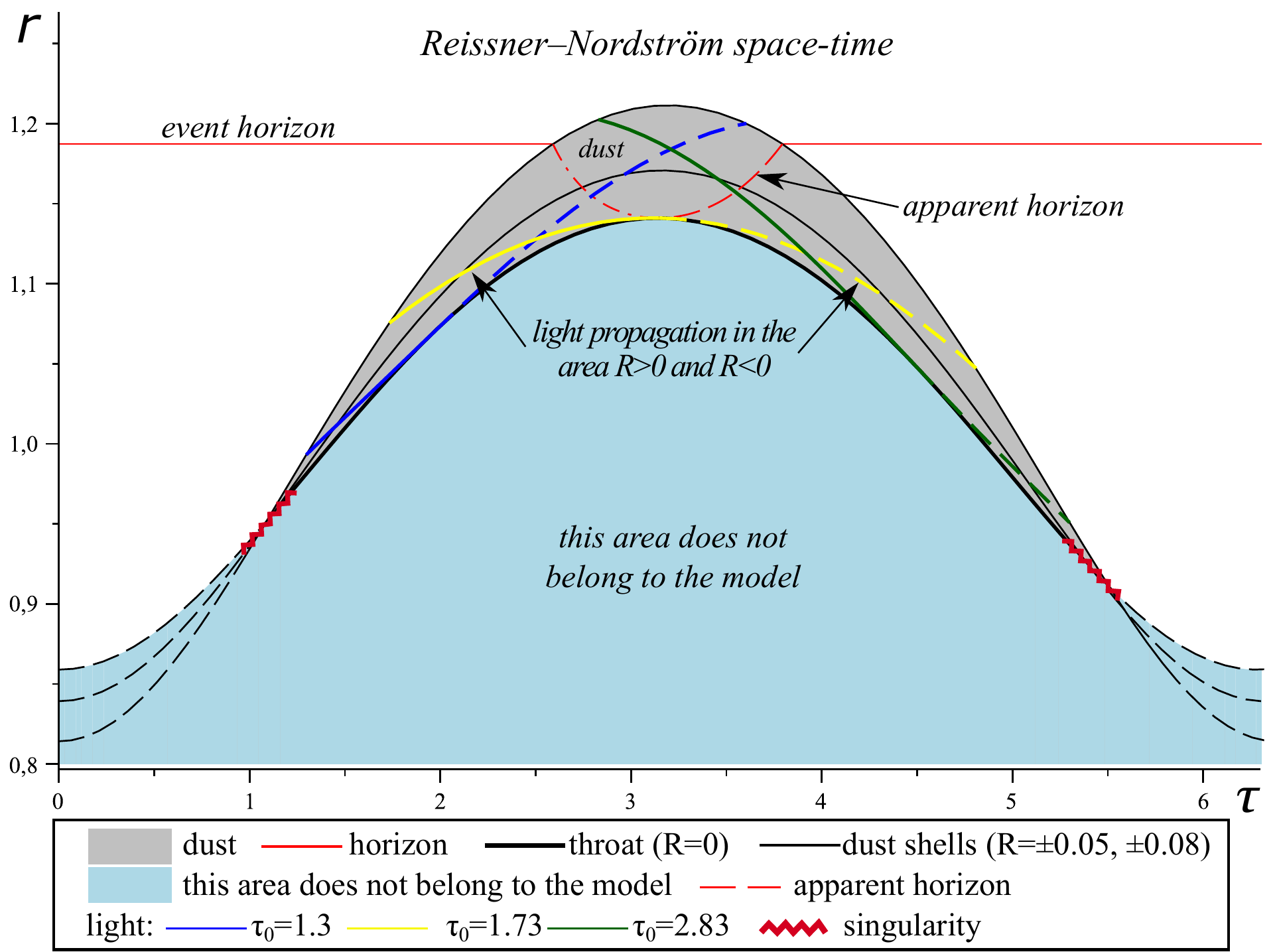}  
\caption{\small
	The figure shows the time dependence $r(R,\tau)$ of dust shells ($R_0=0$, $\pm 0.05$, $\pm 0.08$, 
	$|R|\leq 0.8$) 
	in the case $q=0.99$. The curve $R=0$ depicts the throat; the curve $R=\pm 0.08$ corresponds to
	the outer surface. 
	The blue area below the throat does not belong to the model, it does not describe any part of space-time.
	The dust shells begin and end their motion at singularities, presented by red broken lines.
     The upper unpainted area depicts the \RN\ space-time; the red horizontal line corresponds to the horizon; 
     the red dot-dashed and dashed lines in the dust layer depict the apparent horizon, which is also the location 
     of possible turning points for ingoing and outgoing light rays. The green, yellow, and blue curves correspond to 
     propagation of photons, emitted from the shell $R=0.08$  at different times $\tau_0$, in the region $R>0$ 
     (solid lines) and $R<0$ (dashed lines). Light propagation outside the dust layer is not shown. 
 \label{light_q}}.
\end{figure}  
  
  Equation \rf{r_tau} is the equation of radial motion of a photon in the dust layer, and for our example \rf{ex1} 
  it gives 
\beq                         \label{r_tau2}
      \frac {d r(\tau,R(\tau))}{d\tau} = \frac{1}{\sqrt{1+R^2}}\left(\frac{\Delta\sin\eta}{1-\Delta\cos\eta}-R \right),
\eeq
  where $\eta$ is meant as a function of $\tau$ and $R$, determined by the relations 
\beq  
  	r = (1+R^2)^2 (1-\Delta\cos\eta), \qquad \tau = (1+R^2)^{5/2} (\eta-\Delta\sin\eta).
\eeq  	 
  We consider photons (or light rays) that move radially in the direction from the region $R>0$ to the region $R<0$, 
  which corresponds to the minus sign in the second term in \rf{r_tau}. For motion in the opposite direction, 
  one should replace $R \to -R$ in \eqn{r_tau2}. In the model with $q\neq 0$, the quantity $dr/d\tau$ is always finite; 
  while in the model with pure dust ($q=0$) it diverges at $\eta=0$ and $\eta = 2\pi$; in all cases, a photon crosses 
  the vicinity of the throat $R=0$ in a finite proper time. The results of numerical integration of \eq (\ref{r_tau2}) 
  in the cases $q=0$ and $q\neq 0$ are shown in Figs.\,\ref{light_q0} and \ref{light_q}. 

  In both figures, the gray area depicts the evolution of a dust layer between their initial and final singularities;
  actually, each curve $r(R, \tau)$ in this area, except $R=0$, corresponds to two shells, $R>0$ and $R <0)$.
  The upper white areas in Figs.\,\ref{light_q0} and \ref{light_q} actually correspond to two copies of the outer 
  \Scw\ or \RN\ region, respectively. In both figures one can see photon trajectories that cross the whole dust
   layer in finite proper time $\tau$, which means that the \wh\ parts of these \bh\ space-times are traversable. 

\section{Conclusions}
  
  The main results of this paper may be summarized as follows:
  
\begin{enumerate}
\item 
	It has been shown that the q-Tolman dust clouds can contain \wh\ throats under certain conditions on 
	the arbitrary functions $f(R)$ and $F(R)$ of the general solution to the field equations.   
\item   
   	It has been shown that throats can only exist in the elliptic branch of q-Tolman space-times and are
   	in general located in T-regions. Thus means that if a dust layer is matched to external \RN\ or 
   	Schwarzschild space-time regions, the whole configuration is a black hole rather than a wormhole. 
\item
    The q-Tolman space-times with throats are proven to exist for a finite period of time in their comoving 
    reference frames. If $q=0$ (no \emag\ field), the evolution takes place between two ``spherical'' singularities 
    at which $r \to 0$, while at $q \ne 0$ the initial and final singularities are of shell-sticking nature ($dr/dR \to 0$).
\item
    An analysis of radial null geodesics for particular examples of models under study has shown that the dust
    layers with throats can be traversable in both cases $q=0$ and $q \ne 0$. In other words, a photon can 
    cross such a dust layer more rapidly than this layer collapses. 
\item
    It has been shown that q-Tolman clouds with throats can form traversable \whs\ in closed isotropic cosmological 
    models filled with dust
    --- or, more precisely, connect two copies of such a model. Therefore, each of these universes, 
    taken separately, preserves all its important physical features, including such an issue as a 
    cosmological horizon. However, if the wormhole is large enough, one should also consider the 
    view of ``another universe'' through this wormhole, which needs a separate study. Note also that 
    these wormholes are expand and contracting together with the corresponding cosmological 
    space-time.    
\end{enumerate}     	
 
   Solutions with throats form a subfamily (satisfying the conditions \rf{throat}) in the whole set of 
   q-Tolman solutions. A finite lifetime of models with throats, from one singularity to another, is an important
   feature of this family, while in general, with $q\ne 0$, nonsingular models are possible.
   
   
\Acknow{P.E.K. and S.V.S. are supported by RSF grant No. 21-12-00130. Partially, this work was done in the 
  framework of the Russian Government Program of Competitive Growth of the Kazan Federal University.
  K.B. was supported in part by the RUDN University Strategic Academic Leadership Program, 
  by RFBR Project 19-02-00346, and by the Ministry of Science and Higher Education 
  of the Russian Federation, Project ``Fundamental properties of elementary 
  particles and cosmology'' N 0723-2020-0041}
  

\end{document}